\documentclass[showpacs,showkeys,12pt,preprint,preprintnumbers,nofootinbib,groupedaddress,superscriptaddress,amsmath,amssymb]{revtex4}
\usepackage{graphicx}
\usepackage{amsmath}
\usepackage{color}
\newcommand{\bea}{\begin{eqnarray}}
\newcommand{\ena}{\end{eqnarray}}

\renewcommand{\a}{\alpha}
\renewcommand{\b}{\beta}

\begin{document}
\title{Phase effects from the general neutrino Yukawa matrix on lepton flavor
violation}

\author{Shinya Kanemura}
\email{kanemu@het.phys.sci.osaka-u.ac.jp}
\affiliation{Department of Physics, Osaka University, Toyonaka, Osaka
 560-0043, Japan}

\author{Koichi Matsuda}
\email{matsuda@het.phys.sci.osaka-u.ac.jp}
\affiliation{Department of Physics, Osaka University, Toyonaka, Osaka
 560-0043, Japan}

\author{Toshihiko Ota}
\email{toshi@het.phys.sci.osaka-u.ac.jp}
\affiliation{Department of Physics, Osaka University, Toyonaka, Osaka
 560-0043, Japan}

\author{Tetsuo Shindou}
\email{shindou@sissa.it}
\affiliation{Scuola Internazionale Superiore di Studi Avanzati\\
Via Beirut 2--4, I-34014, Trieste, Italy}

\author{Eiichi Takasugi}
\email{takasugi@het.phys.sci.osaka-u.ac.jp} 
\affiliation{Department of Physics, Osaka University, Toyonaka, Osaka
 560-0043, Japan}

\author{Koji Tsumura}
\email{ko2@het.phys.sci.osaka-u.ac.jp}   
\affiliation{Department of Physics, Osaka University, Toyonaka, Osaka
 560-0043, Japan}

\preprint{OU-HET-525, SISSA 54/2005/EP}

\pacs{12.60.Jv, 14.60.Pq, 14.60.St }

\keywords{Lepton flavor violation, Supersymmetry, Majorana phases}

\begin{abstract}
We examine contributions from Majorana phases to lepton 
flavor violating processes in the framework of the minimal supersymmetric 
standard model with heavy right-handed neutrinos. 
All phases in the complex neutrino Yukawa matrix 
are taken into account in our study. 
We find that in the scenario with universal soft-breaking terms 
sizable phase effects can appear on the lepton flavor 
violating processes such as $\mu \to e \gamma$, $\tau \to e \gamma$, 
and $\tau \to \mu \gamma$.
In particular, the branching ratio of $\mu \to e \gamma$ can be 
considerably enhanced due to the Majorana phases, 
so that it can be much greater than that of $\tau \to \mu \gamma$.
\end{abstract}

\maketitle

\section{Introduction}
In the standard model lepton flavor violation (LFV) is negligible, 
while it can be sizable in new physics models such as 
those based on supersymmetry (SUSY).
Therefore search for LFV can be a good probe of new physics.
Observed tiny neutrino masses may be explained by the seesaw mechanism\cite{seesaw}, 
assuming heavy right-handed Majorana neutrinos, 
which are compatible with the scenario of grand unified theories (GUTs). 
In the framework of SUSY models, LFV is induced through one-loop 
diagrams with slepton mixing\cite{lfv}. 
In the SUSY model with right-handed neutrinos, the slepton mixing 
can be induced from the renormalization group effect of 
the neutrino Yukawa interaction 
between the scale of right-handed neutrino masses and the GUT scale, 
even when soft-SUSY-breaking terms are universal at the GUT scale.

The neutrino mass matrix obtained via the seesaw mechanism generally 
includes two Majorana phases\cite{majorana-cp}. 
They can be directly searched through neutrinoless double beta decays\cite{DKT}.
The existence of these Majorana phases can play an important role 
in various phenomena such as leptogenesis\cite{eps_leptogene}, 
lepton number violating processes and so on.
Searches for these phenomena could provide a hint for 
the neutrino Majorana mass matrix.
Furthermore, as we shall show below, the prediction on LFV can be 
drastically changed by the Majorana phases.

In the present paper, we explore LFV processes such as 
$\mu \to e \gamma$ 
in the framework of the minimal supersymmetric standard model with 
right-handed Majorana neutrinos (MSSMRN) under the assumption of 
universal soft-SUSY-breaking terms at the GUT scale 
$M_{GUT}$. 
Neutrino mass matrix $m_\nu$ is given by 
$m_\nu = Y_\nu^T D_R^{-1} Y_\nu \langle \phi_u^0 \rangle^2$, 
 where $Y_\nu$ is the neutrino Yukawa matrix, 
$D_R$ is the right-handed neutrino mass matrix which is diagonal, 
and $\phi_u^0$ is the neutral component of the Higgs doublet 
with hypercharge $-1/2$. 
In the basis where the charged lepton mass matrix is diagonal, 
the neutrino Dirac mass matrix 
$m_D \equiv Y_\nu \langle \phi_u^0 \rangle$ can 
be parameterized by\cite{Casas:2001sr,Pascoli:2003rq}
\begin{align}
m_D = \sqrt{D_R} R \sqrt{D_\nu} U^\dag \; ,
\label{mD}
\end{align}
where 
$D_\nu$ is the eigenmatrix of neutrino masses, 
$R$ is a complex orthogonal matrix ($R^T R = R R^T = {\bf 1}$),
and $U$ is the neutrino mixing matrix. 
In Refs.~\cite{Pascoli:2003rq,Deppisch:2002vz}, the decay 
rates of $\ell_i \to \ell_j \gamma~(i \ne j)$ are evaluated by assuming that 
$R$ is a real orthogonal matrix 
and that the right-handed neutrino masses are degenerate; i.e., 
$D_R = M \times {\bf 1}$ 
where $M$ is the heavy Majorana mass scale. 
Under this assumption, the effect of Majorana phases on the 
low energy phenomena is screened. 
The relation among the branching ratios is given by
\begin{align}
{\rm Br}(\mu\to e \gamma) \simeq \frac{{\rm Br}(\tau\to e \gamma)}
{{\rm Br}(\tau\to \bar{\nu}_e \nu_\tau e)}
 \ll {\rm Br}(\tau \to \mu \gamma)\;,
\end{align}
where current neutrino data have been used.
The hierarchical $D_R$ case 
with a real $R$ has been analyzed in Ref.~\cite{Casas:2001sr}. 
On the other hand, the importance of the treatment of $R$ 
as a complex matrix has been pointed out in Ref.~\cite{Pascoli:2003rq},
by showing that phases in $R$ can give a substantial 
effect on low energy phenomena. 

In this paper, we discuss the role of the imaginary part of $R$, 
and study the combined effect with Majorana phases in neutrino 
mixing matrix on the branching ratios of the LFV processes. 
We assume that $D_R = M \times {\bf 1}$. 
We obtain analytic expressions of the branching ratios in 
two limiting cases: i.e., 
one is the case with $R$ being approximately a real orthogonal matrix, 
and the other is with $R$ being a typical complex matrix.
We find that
\begin{align}
{\rm Br}(\mu\to e \gamma) \simeq \frac{{\rm Br}(\tau\to e \gamma)}
{{\rm Br}(\tau\to \bar{\nu}_e \nu_\tau e)}
 \gg {\rm Br}(\tau \to \mu \gamma)\;,
\end{align}
in the wide range of the parameter space for a typical complex matrix $R$. 
The branching ratio of $\mu \to e \gamma$ can be enhanced 
in comparison with that of $\tau \to \mu \gamma$.
This is a novel feature with a complex $R$. 
We also give numerical calculations in order to see how 
these two limiting cases are extrapolated. 

\section{Evaluation of LFV branching ratios}
In this section, we briefly review LFV in the MSSMRN, and
discuss the Majorana phase effects on LFV processes.

In the model based on SUSY, LFV processes can occur 
at the low energy scale through the slepton mixing. 
In the MSSMRN, sizable off-diagonal elements of the slepton mass matrix can be induced  by renormalization 
group effects due to the neutrino 
Yukawa interaction between $M_{GUT}$ and $M$, even when
universal soft-breaking masses are assumed at $M_{GUT}$. 
The induced off-diagonal elements are 
approximately expressed as\cite{lfv}
\begin{align}
(m_{\tilde{L}}^2)_{ij}\simeq \frac{6 m_0^2 + |A_0|^2}{16 \pi^2}
\ln \frac{M_{GUT}}{M}(Y_\nu^\dag Y_\nu)_{ij} \quad (i \ne j)\;,
\end{align}
where $m_0$ and $A_0$ are universal soft-SUSY-breaking parameters. 
The decay rates for LFV processes $\ell_i \to \ell_j \gamma~(i \ne j)$ 
are given by
\begin{align}
\Gamma(\ell_i \to \ell_j \gamma) \simeq 
\frac{\alpha_{EM}^3 m^5_{\ell_i}}{192 \pi^3} 
\frac{|(m_{\tilde{L}}^2)_{ij}|^2}{m_{SUSY}^8} \tan^2 \beta \; ,
\end{align}
where $\alpha_{EM}$ is the fine structure constant, $m_{SUSY}^{}$ 
represents the typical mass scale of SUSY particles, and 
$\tan \beta$ is the ratio of vacuum expectation values of 
the two Higgs doublets. 
The branching ratios are related to each other as 
\begin{align}
\frac{{\rm Br}(\mu \to e \gamma)}{{\rm Br}(\tau \to \mu \gamma)} 
&\simeq  \frac{1}{{\rm Br}(\tau\to \bar{\nu}_e \nu_\tau e)} 
\frac{|(m_D^{\dag}m_D^{})_{12}|^2}
{|(m_D^{\dag}m_D^{})_{23}|^2} 
\sim 5.6 \times
\frac{|(m_D^{\dag}m_D^{})_{12}|^2}
{|(m_D^{\dag}m_D^{})_{23}|^2}\;,
\nonumber\\
\frac{{\rm Br}(\tau \to e \gamma)}{{\rm Br}(\tau \to \mu \gamma)} 
&\simeq \frac{|(m_D^{\dag}m_D^{})_{13}|^2}
{|(m_D^{\dag}m_D^{})_{23}|^2}
\;,
\label{RatioBr}
\end{align}
where experimental result ${\rm Br}(\tau\to \bar{\nu}_e \nu_\tau e)=0.1784$ 
is used. These ratios are determined only by the neutrino Yukawa matrix.

We work on the basis that the right-handed neutrino mass matrix
is diagonal, and assume that the matrix is approximately proportional to 
the identity matrix; i.e. 
$D_R  \simeq M\times {\bf 1}$. By using Eq.~\eqref{mD} we obtain
\begin{align}
m_D^{\dag}m_D^{}
\simeq M U\sqrt{D_\nu} 
R^\dag R \sqrt{D_\nu}U^{\dag}
= M U\sqrt{D_\nu} 
 Q^\dag Q \sqrt{D_\nu}U^{\dag}\;.
\end{align}
Here we have introduced a real orthogonal matrix 
$O$ by $R=OQ$, where $Q$ is a product of $Q_a(a=1$-$3)$ with
\begin{align}
Q_1=\begin{pmatrix}\cosh y_1&i\sinh y_1&0 \\
	           -i\sinh y_1&\cosh y_1&0 \\
	           0&0&1 \end{pmatrix}\;,
Q_2=\begin{pmatrix}1&0&0 \\
		0& \cosh y_2&i\sinh y_2 \\
	           0&-i\sinh y_2&\cosh y_2 \end{pmatrix}\;,
Q_3=\begin{pmatrix}\cosh y_3&0&i\sinh y_3 \\
		0&1&0 \\
	           -i\sinh y_3&0&\cosh y_3 \end{pmatrix}\;.
\end{align}
The matrices $Q_a$ satisfy that $Q_a^\dag =Q_a$ and 
$Q_a^2(y_a)= Q_a(2y_a)$. 
The matrix $Q$ plays a role not only to introduce the complex phases 
but also to change the size of Yukawa couplings.\footnote{
We note that fine tuning of order $\mathcal{O}(e^{y_a})$ is necessary
to obtain the light neutrino mass scale in the case of $y_a>1$.}

The neutrino mixing matrix $U$ is separated into 
two parts, $U=U_{MNS} P$,
where $U_{MNS}$ is the Maki-Nakagawa-Sakata matrix\cite{mns} in
the phase convention of Ref.~\cite{PDG-phase} and 
$P$ is the Majorana phase matrix given by
$P={\rm diag}(1,e^{i\a_0},e^{i\b_0})$
with $\a_0$ and $\b_0$ being Majorana CP violation phases
\cite{majorana-cp}. 
In order to see qualitative features, we here take 
the Bi-maximal mixing solution\cite{BiMax} 
\begin{align}
U_{MNS}^{Bi-max}=\begin{pmatrix} \frac{1}{\sqrt{2}} &\frac{1}{\sqrt{2}} &0\\ 
-\frac12&\frac12&\frac{1}{\sqrt{2}}\\
\frac12&-\frac12&\frac{1}{\sqrt{2}}\\
\end{pmatrix}
\;,
\end{align}
for analytic calculations.
In particular, we consider the following three cases for $D_\nu$;
the normal hierarchical (NH) case ($m_1 \ll m_2 \ll m_3$), 
the inverse hierarchical (IH) case ($m_3 \ll m_1 \sim m_2 $),
and the quasi-degenerate (QD) case ($m_1 \simeq m_2 \sim m_3$);
\begin{align}
&{\rm NH:} \quad m_1 \simeq 0 \;, \ m_2 \simeq \sqrt{\Delta m^2_\odot} 
 \;,  \
 m_3 \simeq \sqrt{\Delta m^2_{\rm atm}}\;,\\
&{\rm IH:} \quad m_1 \simeq \sqrt{\Delta m^2_{\rm atm}} \left( 1 - \frac{1}{2} \frac{\Delta
 m^2_\odot}{\Delta m^2_{\rm atm}} \right) \;, \
 m_2 \simeq \sqrt{\Delta m^2_{\rm atm}} \;, \
 m_3 \simeq 0\;,\\
&{\rm QD:} \quad m_1 \equiv m \;, \
 m_2 \simeq m + \frac{\Delta m^2_\odot}{2m}\;, \
 m_3 \simeq m + \frac{\Delta m^2_{\rm atm}}{2m}\;. \
\end{align}
Here,  $\Delta m^2_\odot \equiv m_2^2-m_1^2 (=8.0 \times 10^{-5}{\rm eV}^2)$\cite{Miknaitis:2005rw}
is the squared mass difference for the solar neutrino mixing, and 
$\Delta m^2_{\rm atm} \equiv |m_3^2-m_2^2| (= 2.5 \times 10^{-3}{\rm eV}^2)$\cite{Moriyama:2005yr}
is that for the atmospheric neutrino mixing. 

To evaluate $m_D^\dag m_D^{}$, we consider the following two limiting cases. 
\begin{enumerate}
\item[(a)] The small $y_a$ limit ($R$ is real.)~:\\
We have $Q={\bf 1}$, and thus 
$m_D^{\dag}m_D^{}=MU_{MNS}^{Bi-max}D_\nu {U_{MNS}^{Bi-max}}^\dag$, 
where the elements of $m_D^{\dag}m_D^{}$ 
are determined by neutrino masses and the mixing matrix as
\begin{align}
	(m_D^{\dag}m_D^{})_{12}&=(m_D^{\dag}m_D^{})_{13}
	=-\frac{M}{2\sqrt 2}(m_2-m_1)\;, \nonumber\\
	(m_D^{\dag}m_D^{})_{23}&= \frac{M}4 (m_1+m_2-2m_3)\;.
\end{align}
We then obtain from Eq.~\eqref{RatioBr} that
\begin{align}
\frac{{\rm Br}(\mu \to e \gamma)}{{\rm Br}(\tau \to \mu \gamma)} 
\simeq \left\{\begin{array}{ccc} 
 5.6 \times \frac{1}{2} \left(\frac{\Delta m_\odot^2}{\Delta m_{\rm atm}^2}
 \right)&\simeq 0.23& {\rm for\;\; NH}\\
 5.6 \times \frac{1}{8} \left( \frac{\Delta m_\odot^2}{\Delta m_{\rm atm}^2}
 \right)^2&\simeq 7.7 \times 10^{-4} &{\rm for \;\;IH}\\
 5.6 \times \frac{1}{2} \left( \frac{\Delta m_\odot^2}{\Delta m_{\rm atm}^2}
 \right)^2&\simeq  3.1 \times 10^{-3}& {\rm for\;\; QD}\end{array}
 \right. \;.
\end{align} 
For all the cases, it turns out that ${\rm Br}(\mu \to e \gamma)\simeq 
5.6 \times {\rm Br}(\tau \to e \gamma) \ll {\rm Br}(\tau \to \mu \gamma)$, 
as pointed out in Refs.~\cite{Pascoli:2003rq,Deppisch:2002vz}. 
In this limit, the Majorana phases do not affect the LFV processes.
	
\item[(b)] The large $y_a$ case~:\\
The matrix $Q$ has a simple form. 
First, the matrices $Q_a$ behave as
\begin{align}
Q_a \simeq \frac{e^{y_a}}{\sqrt{2}} {\cal Q}_a\;,
\end{align}
where
\begin{align}
{\cal Q}_1=\frac{1}{\sqrt{2}} \begin{pmatrix}1 &i &0\\ -i &1 &0\\ 
 0&0&0 \end{pmatrix}\;,
{\cal Q}_2=\frac{1}{\sqrt{2}} \begin{pmatrix}0 &0 &0\\ 0 &1 &i\\ 
 0&-i&1 \end{pmatrix}\;,
{\cal Q}_3=\frac{1}{\sqrt{2}} \begin{pmatrix}1 &0 &i\\ 0 &0 &0\\ 
 -i&0&1 \end{pmatrix}\;.
\end{align}
They satisfy
${\cal Q}_a^\dag={\cal Q}_a$ and ${\cal Q}_a^2=\sqrt{2} {\cal Q}_a$. 
As for the product of ${\cal Q}_a$ such as ${\cal Q} \in \{ {\cal Q}_a, 
{\cal Q}_b {\cal Q}_a, {\cal Q}_c {\cal Q}_b {\cal Q}_a \}$,
we find an interesting relation as
\begin{align}
{\cal Q}^\dag {\cal Q}=\sqrt{2} {\cal Q}_a\;.
\label{qq}
\end{align}
By using Eq.~\eqref{qq} $Q^\dag Q$ is expressed by
\begin{align}
Q^\dag Q \simeq \frac{e^{2(y_1+y_2+y_3)}}{4\sqrt{2}} {\cal Q}_a\;.
\end{align}
This means that $Q^\dag Q$ is characterized by three independent 
matrices ${\cal Q}_a$($a=1$-$3$) for large $y_a$.
Thus, we examine the following three cases, taking $R=OQ_a\simeq e^{y_a}O {\cal Q}_a/\sqrt{2}$.
\begin{enumerate}
\item[(b-1)] $R=OQ_1$\\
We have
\begin{align}
(m_D^{\dag}m_D^{})_{12} &= - (m_D^{\dag}m_D^{})_{13} =
\left( \frac{Me^{2y_1}}{2}\right)
\frac{m_2 - m_1 + i 2 {\sqrt{m_1 m_2}} \cos \alpha_0}{2 \sqrt{2}}\;,
\nonumber\\
(m_D^{\dag}m_D^{})_{23}&= - \left( \frac{Me^{2y_1}}{2}\right)
\frac{m_1 + m_2 - 2  \sqrt{m_1 m_2} \sin \alpha_0}{4}\;.
\end{align}
Thus the LFV branching ratios are related as  
${\rm Br}(\mu \to e \gamma) \simeq 5.6 \times {\rm Br}(\tau \to e \gamma)$ 
for all cases.  
For the NH case, we obtain
${\rm Br}(\mu \to e \gamma) \simeq 11.2 \times {\rm Br}(\tau \to \mu \gamma)$. 
For the IH and the QD cases, one finds
\begin{align}
\frac{{\rm Br}(\mu \to e \gamma)}{{\rm Br}(\tau \to \mu \gamma)} 
\simeq 11.2 \times \frac{\cos^2 \alpha_0}{(1+\sin \alpha_0)^2}\;.
\end{align}
This ratio is a function of $\alpha_0$. It is 11.2 for $\alpha_0=0$ or $\pi$, 
and 0 for $\alpha_0=\pi/2$. 

\item[(b-2)]$R=OQ_2$\\
The difference of the Majorana phases $\alpha_0-\beta_0$ enters into 
$m_D^\dag m_D^{}$,
\begin{align}
(m_D^{\dag}m_D^{})_{12}& =\left(\frac{Me^{2y_2}}{2}\right)
\frac{m_2 + i \sqrt{2} \sqrt{m_2 m_3} 
e^{i \left( \a_0 - \b_0  \right)}}{2 \sqrt{2}}\;,
\nonumber\\
(m_D^{\dag}m_D^{})_{13}&=\left(\frac{Me^{2y_2}}{2}\right)
\frac{-m_2 + i \sqrt{2} \sqrt{m_2 m_3} 
e^{i \left( \a_0  - \b_0  \right)}}{2 \sqrt{2}}\;,
\nonumber\\
(m_D^{\dag}m_D^{})_{23}&=\left(\frac{Me^{2y_2}}{2}\right)
\frac{ - m_2 + 2\,m_3 + 2\sqrt2 i \sqrt{m_2 m_3} \cos (\a_0-\b_0)}{4}\;.
\end{align}
For the NH case and the IH case, the branching ratios of $\ell_i \to \ell_j \gamma$ are related to each other as 
${\rm Br}(\mu \to e \gamma) \simeq 5.6 \times {\rm Br}(\tau \to e \gamma)$, and
\begin{align}
\frac{{\rm Br}(\mu \to e \gamma)}{{\rm Br}(\tau \to \mu \gamma)} 
\simeq \left\{\begin{array}{cc}5.6 \times \sqrt{\frac{\Delta m_\odot^2}{\Delta m_{\rm atm}^2}}
\simeq 1.0& {\rm for\;\; NH}\\
11.2 &{\rm for \;\;IH}\end{array}
 \right. .
\end{align}
For the QD case, we obtain
\begin{align}
\frac{{\rm Br}(\mu \to e \gamma)}{{\rm Br}(\tau \to \mu \gamma)} 
&\simeq 11.2 \times \frac{3+2\sqrt2 \sin(\a_0-\b_0)}{1+8\cos^2(\a_0-\b_0)}\;,
\nonumber\\
\frac{{\rm Br}(\tau \to e \gamma)}{{\rm Br}(\tau \to \mu \gamma)} 
&\simeq \frac{2(3-2\sqrt2 \sin(\a_0-\b_0))}{1+8\cos^2(\a_0-\b_0)}\;.
\label{eq:QD23Br}
\end{align}
We have ${\rm Br}(\mu \to e \gamma)\simeq5.6 \times {\rm Br}(\tau \to e \gamma)\simeq3.7 \times {\rm Br}(\tau \to \mu \gamma)$ 
for $\a_0-\b_0=0$ or $\pi$.
The ratio ${\rm Br}(\mu \to e \gamma)/{\rm Br}(\tau \to \mu \gamma)$ 
takes its minimum value 1.9 at  $\a_0-\b_0 \simeq -\pi/2$.  

\item[(b-3)]$R=OQ_3$	\\
The Majorana phase $\beta_0$ enters into 
$m_D^\dag m_D^{}$. We obtain 
\begin{align}
(m_D^{\dag}m_D^{})_{12} &=\left(\frac{Me^{2y_{3}}}{2}\right)
\frac{ - m_1 + i \sqrt{2} \sqrt{m_1 m_3} e^{-i \,\beta_0 }}{2 \sqrt{2}}\;,
\nonumber\\
(m_D^{\dag}m_D^{})_{13}&=\left(\frac{Me^{2y_{3}}}{2}\right)
\frac{m_1 + i \sqrt{2} \sqrt{m_1 m_3} e^{-i \,\beta_0 }}{2 \sqrt{2}}\;, 
\nonumber\\ 
(m_D^{\dag}m_D^{})_{23}&=\left(\frac{Me^{2y_{3}}}{2}\right)
\frac{-m_1 + 2 m_3 - i 2 \sqrt{2} \sqrt{m_1 m_3} \cos \beta_0}{4}\;.
\end{align}
In this case $\left|(m_D^{\dag}m_D^{})_{ij}\right|^2$ 
can be obtained from case (b-2)
by replacing $m_2$ with $m_1$ and $\alpha_0-\beta_0$ with $\pi-\beta_0$.
The branching ratio ${\rm Br}(\mu \to e \gamma)$ is suppressed 
in the NH case because of ${\rm Br}(\mu \to e \gamma)
\simeq 5.6 \times {\rm Br}(\tau \to e \gamma)\simeq 
5.6 \times (m_1^2/\Delta m_{\rm atm}^2) {\rm Br}(\tau \to \mu \gamma) \ll {\rm Br}(\tau \to \mu \gamma)$. 
For the IH case, the branching ratios are related to each other as 
${\rm Br}(\mu \to e \gamma) \simeq 5.6 \times{\rm Br}(\tau \to e \gamma) \simeq 11.2 \times {\rm Br}(\tau \to \mu \gamma)$. 
For the QD case, relation among the ratios of branching ratios is obtained 
from Eq.\eqref{eq:QD23Br} by changing $\alpha_0-\beta_0$  to $\pi-\beta_0$.
\end{enumerate}
The results are summarized in Table 1. 
\end{enumerate}

\begin{table}
\begin{tabular}{|c|c|c|c|c|c|}
\hline
\multicolumn{2}{|c|}{}&\multicolumn{1}{|c|}{small $y_a$} 
& \multicolumn{3}{|c|}{large $y_a$} \\ 
\cline{3-6}
\multicolumn{2}{|c|}{}&
\multicolumn{1}{|c|}{$Q={\bf 1}$} & 
\multicolumn{1}{|c|}{$Q_1$} &
\multicolumn{1}{|c|}{$Q_2$ }&
\multicolumn{1}{|c|}{$Q_3$} \\
\hline
$\frac{{\rm Br}(\mu \to e \gamma)}{{\rm Br}(\tau \to \mu \gamma)}$
&NH&$0.23$&11.2& $1.0$ &$\ll 1$\\
\cline{2-6}
 &IH& $7.7\times 10^{-4}$ & $11.2 \times \frac{\cos^2\a_0}{(1+\sin\a_0)^2}$
 &11.2&11.2\\
\cline{2-3}\cline{5-6}
 &QD& $3.1\times 10^{-3}$ &
 &$11.2 \times \frac{3-2\sqrt2 \sin(\a_0-\b_0)}{1+8\cos^2(\a_0-\b_0)}$
 &$11.2 \times \frac{3+2\sqrt2 \sin\b_0}{1+8\cos^2\b_0}$\\
\hline
$\frac{{\rm Br}(\mu \to e \gamma)}{{\rm Br}(\tau \to e \gamma)}$
&NH&5.6&5.6&5.6&5.6\\
\cline{2-2}
 &IH&&&&\\
\cline{2-2}\cline{5-6}
 &QD&&&$5.6 \times \frac{3-2\sqrt2 \sin(\a_0-\b_0)}{3+2\sqrt2 \sin(\a_0-\b_0)}$
 &$5.6 \times \frac{3+2\sqrt2 \sin\b_0}{3-2\sqrt2 \sin\b_0}$\\
\hline
\end{tabular}
\caption{Summary of the ratios of the LFV processes.}
\end{table}

For the ratio ${\rm Br}(\mu \to e \gamma)/{\rm Br}(\tau \to e \gamma)$, 
$Q$ does not contribute except for the QD case 
with $Q=Q_2$ or $Q_3$ where the Majorana phases give a significant effect.
The drastic effect occurs for 
${\rm Br}(\mu \to e \gamma)/{\rm Br}(\tau \to \mu \gamma)$ 
by $Q$ or by the interplay between $Q$ and the
Majorana phases. The substantial enhancement arises in
${\rm Br}(\mu \to e \gamma)$, which is a quite different feature
from the case with $Q={\bf 1}$. 
By the introduction of $Q(\ne {\bf 1})$, the Majorana phases can affect the
LFV processes. This fact is a quite 
interesting because the observation of the LFV 
processes would give useful information of Majorana phases. 

\section{Numerical results}
In the previous section, we consider the two limiting cases for 
the parameter $y_a$. When $y_a$ take the intermediate values, 
we may guess the result by extrapolating from the two limits, but 
some non-trivial structure might appear. Therefore, 
we perform the numerical evaluation of the LFV branching ratios
 for three typical cases, $R=OQ_a(a=1$-$3)$. 
Neutrino mixing parameters are taken to be $\tan^2\theta_\odot = 0.45$\cite{Miknaitis:2005rw}, $\sin2\theta_{atm}=1$\cite{Moriyama:2005yr}, and $\sin\theta_{13}=0$. 
The values for $M$ and $M_{GUT}$ are taken as $M=10^{10}$GeV and 
$M_{GUT}=2 \times 10^{16}$GeV.
The SUSY parameters are taken to be $m_0=A_0=m_{SUSY}^{}=100$GeV 
and $\tan \beta=10$. For standard model parameters $\a_{EM}^{}=1/137$ and $v=246$GeV are used. It will be shown that the ratios of the branching ratios
are not sensitive to SUSY parameters, right-handed neutrino mass scale, and the GUT scale.

We analyze the $y_a$ dependences of 
${\rm Br}(\mu \to e \gamma)/{\rm Br}(\tau \to \mu \gamma)$.
The result for the NH case is shown in Fig.~1. 
We find the smooth extrapolation in 
${\rm Br}(\mu \to e \gamma)/{\rm Br}(\tau \to \mu \gamma)$
between ${\cal O}(0.1)$ and ${\cal O}(1)$ for $R=O Q_2$ with $\alpha_0-\beta_0=0$ and between ${\cal O}(0.1)$ to ${\cal O}(10^{-6}) \ll 1$ for $R=O Q_3$.  For  $R=O Q_1$, some structure is observed 
between ${\cal O}(0.1)$ and ${\cal O}(10)$. The ratio blows up around 
$y \sim 1.3$ due to the vanishing ${\rm Br}(\tau \to \mu \gamma)$. 
There is no $\a_0$ dependence. 
\begin{figure}
\begin{center}
\setlength{\unitlength}{1mm}
\begin{picture}(100,55)
\put(5,2){\includegraphics[width=8cm]{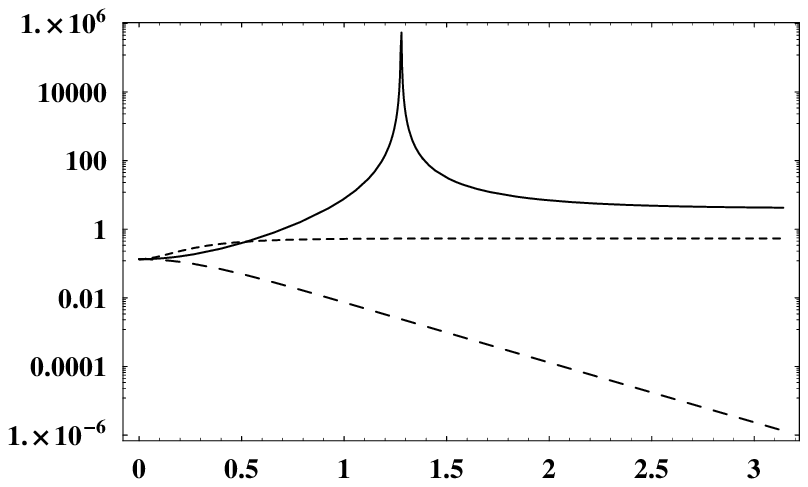}}
\put(50,-1){$y_a$}
\put(0,10){\rotatebox{90}{${\rm Br}(\mu \to e \gamma)/{\rm Br}(\tau \to \mu \gamma)$}}
\end{picture}
\caption{The ratio of the branching ratios is shown in the NH case for $Q_{1}$ (solid curve), for $Q_{2}$ with $\alpha_0-\beta_0=0$ (dotted curve), and for $Q_{3}$ (dashed curve).}
\end{center}
\end{figure}

The IH case is shown in Fig.~2. 
The dotted (dashed) curve represents ${\rm Br}(\mu \to e \gamma)/{\rm Br}(\tau \to \mu \gamma)$ 
for $R=O Q_2$ ($O Q_3$) where the 
smooth extrapolation is found between very small value to 
about $50$ ($2$), where there is no $\a_0$ dependence. 
The case $R=O Q_1$ is shown for solid curves, which has 
the Majorana phase $\a_0$ dependence.
For all cases, we find the smooth extrapolations between two 
limiting values, the small $y_a$ and the large $y_a$. 
\begin{figure}
\begin{center}
\setlength{\unitlength}{1mm}
\begin{picture}(100,55)
\put(5,2){\includegraphics[width=8cm]{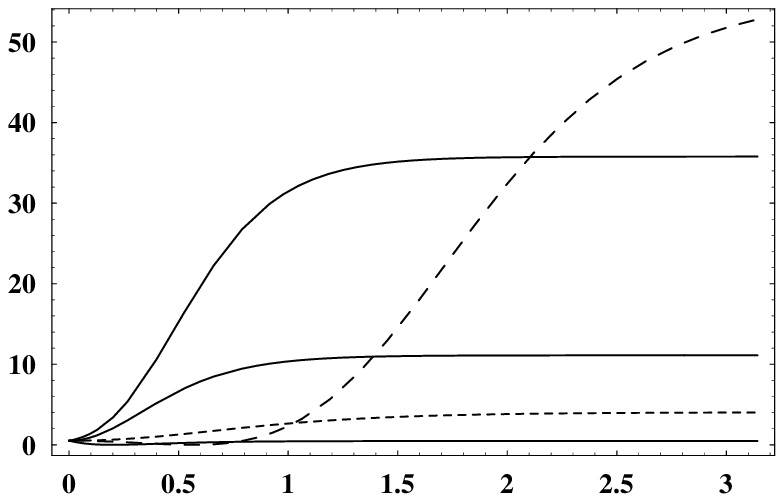}}
\put(46,-1){$y_a$}
\put(67,37.8){{\scriptsize $\alpha_0=3\pi/4$}}
\put(68,17.8){{\scriptsize $\alpha_0=0$}}
\put(67,9){{\scriptsize $\alpha_0=3\pi/2$}}
\put(0,10.2){\rotatebox{90}{${\rm Br}(\mu \to e \gamma)/{\rm Br}(\tau \to \mu \gamma)$}}
\end{picture}
\caption{The ratio of the branching ratios is shown in the IH case for $Q_1$ with $\alpha_0 = 0, 3 \pi/4, 3 \pi/2$ (solid curve), for $Q_2$ (dotted curve), and for $Q_3$ (dashed curve).}
\end{center}
\end{figure}

For the QD case with $R=O Q_2$, the ratio of the branching ratios 
depends on 
$\a_0-\b_0$, and is roughly obtained by replacing $\b_0$ to 
$\pi - (\a_0-\beta_0)$ in the formula for $R=O Q_3$. 
The results for $R=O Q_1$ are similar to those for the IH case with 
$R=O Q_1$. 
In Fig.~3, we show the $y_3$ dependence for the case with $R=O Q_3$ 
for $\a_0-\b_0=0, 3 \pi/4, 3 \pi/2$. 
The enhancement occurs for $\a_0-\b_0=3 \pi/4$ because 
${\rm Br}(\tau \to \mu \gamma)$ is suppressed. 
\begin{figure}
\begin{center}
\setlength{\unitlength}{1mm}
\begin{picture}(100,55)
\put(5,2){\includegraphics[width=8cm]{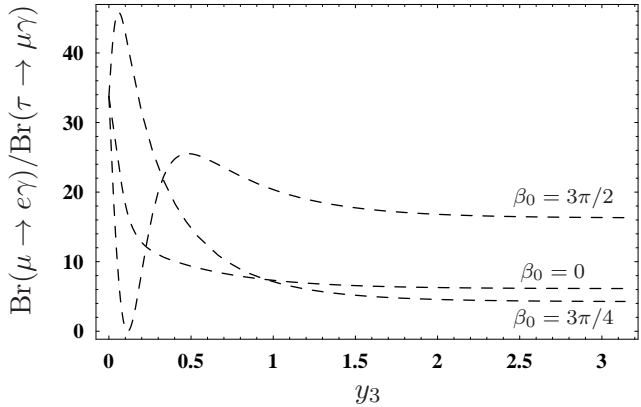}}
\put(46,-1){$y_3$}
\put(67,25){{\scriptsize $\beta_0=3\pi/2$}}
\put(68,15){{\scriptsize $\beta_0=0$}}
\put(67,8.8){{\scriptsize $\beta_0=3\pi/4$}}
\put(0,10){\rotatebox{90}{${\rm Br}(\mu \to e \gamma)/{\rm Br}(\tau \to \mu \gamma)$}}
\end{picture}
\caption{The ratio of the branching ratios is shown in the QD case for $Q_{3}$ with $\beta_0 = 0, 3 \pi/4, 3 \pi/2$.}
\end{center}
\end{figure}

In Fig.~4, ${\rm Br}(\mu \to e \gamma)$ 
with $R=O Q_1$ is shown as a function of $y_1$. 
As $y_1$ grows, the neutrino Yukawa couplings become large 
for all the neutrino mass spectrum. 
Thus, the smooth extrapolation is obtained,
so that the two limiting cases give 
the general trend of the $y_a$ dependence. 
In many cases, the Majorana phases play an important role 
on the prediction of the LFV processes. 
Therefore, we can obtain useful information of the Majorana phases 
from the experimental data of the LFV processes. 
\begin{figure}
\begin{center}
\setlength{\unitlength}{1mm}
\begin{picture}(100,55)
\put(5,2){\includegraphics[width=8cm]{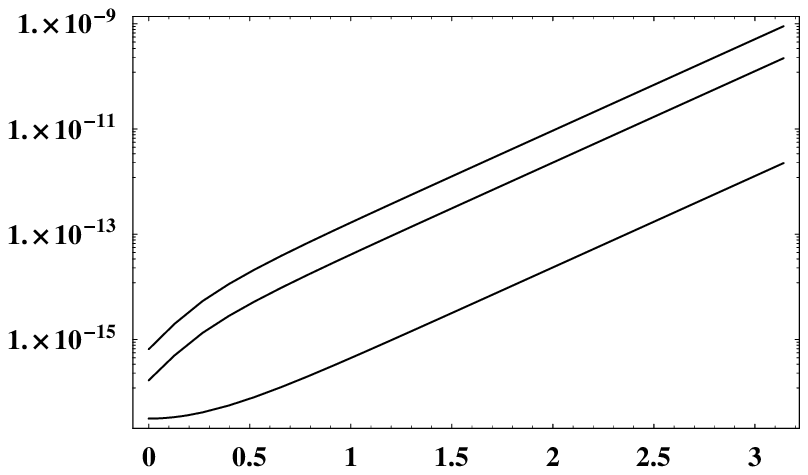}}
\put(50,-1){$y_1$}
\put(60,20){NH}
\put(50,26){\rotatebox{25}{IH {\scriptsize with $\alpha_0=3\pi/4$}}}
\put(50,35){\rotatebox{25}{QD {\scriptsize with $\alpha_0=3\pi/4$}}}
\put(0,20){\rotatebox{90}{${\rm Br}(\mu \to e \gamma)$}}
\end{picture}
\caption{The LFV branching ratios ${\rm Br}(\mu \to e \gamma)$ 
are shown in the NH case, the IH case with $\alpha_0=3\pi/4$ 
and the QD case with $\alpha_0=3\pi/4$ for $R=O Q_1$. }
\end{center}
\end{figure}

\section{Conclusion}
We have shown the importance of the complex nature of the neutrino 
Yukawa matrix for the case of the degenerate right-handed neutrino 
masses. With the complex $R$, the Majorana 
phases play an important role for the prediction of the LFV processes.
In order to see the effect analytically, 
we have taken the parameterization, $R=OQ$.  
We have considered the two limiting cases; 
the small $y_a$ case with $Q = {\bf 1}$ and 
the large $y_a$ case with complex matrix $Q$. 
We have obtained the analytic expressions for ratios of 
the branching ratios of $\mu \to e \gamma$, $\tau \to \mu \gamma$ 
and $\tau \to e \gamma$, which are shown in Table 1. 
The effect of $Q$ is sizable and gives enhancement of 
${\rm Br}(\mu \to e \gamma)/{\rm Br}(\tau \to \mu \gamma)$ 
in many cases. 
In particular, the Majorana phases contribute to some cases. 
This would give a possibility to obtain useful information of 
Majorana phases by observing the LFV processes. 
This is quite interesting and important because
extracting the information for the Majorana phases 
can be used to examine the nature of neutrinos.

It may also be interesting to discuss the possibility to determine 
the neutrino Yukawa matrix by analysing the double beta decay, 
the $\mu^- \to e^+$ \cite{DKT,LTY} and $\mu^- \to \mu^+$ 
conversion \cite{ET}, the LFV processes which occur through SUSY 
contributions, and the leptogenesis. 

\vskip 5mm
\noindent {\large {\bf Acknowledgment}}\\
This work is supported in part by 
the Japanese Grant-in-Aid for Scientific Research of
Ministry of Education, Science, Sports and Culture, Government of Japan, 
Nos.12047218 and 17043008, and also by Japan Society for Promotion of 
Science (Nos. 15-03693 and 15-03700).


\end{document}